\documentclass[twocolumn,aps,prc,showpacs,superscriptaddress]{revtex4}
\usepackage{epsfig,graphics}
\usepackage{caption}
\usepackage{subfigure}
\usepackage{bm}% bold math
\usepackage{amsmath}
\usepackage[usenames]{color}
\usepackage[colorlinks,citecolor=blue]{hyperref}

\begin{document}

\title{Study on higher moments of net-charge multiplicity distributions using a multiphase transport model}

\author{Ling Huang}
\affiliation{Shanghai Institute of Applied Physics, Chinese Academy of Sciences, Shanghai 201800, China}
\affiliation{University of Chinese Academy of Sciences, Beijing 100049, China}
\affiliation{Key Laboratory of Nuclear Physics and Ion-beam Application (MOE), Institute of Modern Physics, Fudan University, Shanghai 200433, China}

\author{Guo-Liang Ma}
\email[]{glma@fudan.edu.cn}
\affiliation{Key Laboratory of Nuclear Physics and Ion-beam Application (MOE), Institute of Modern Physics, Fudan University, Shanghai 200433, China}
\affiliation{Shanghai Institute of Applied Physics, Chinese Academy of Sciences, Shanghai 201800, China}

%\date{\today}

\begin{abstract}
The moments and moment products of conserved charges are believed to be sensitive to critical fluctuations, which have been adopted in determining the QCD critical point. Using a dynamical multiphase transport model, we reproduce the centrality and energy dependences of moments and moment products of net-charge multiplicity distributions in Au+Au collisions measured by the Beam Energy Scan program at the RHIC. No non-monotonic energy dependence is observed. We infer that the moment products develop during the dynamical evolution of heavy-ion collisions. The observed difference based on the expectation of the Poisson baseline indicates a positive two-particle correlation between positively and negatively charged particles, which can arise from different dynamical processes at different stages. Therefore,  to adopt moments and moment products of net-charge multiplicity distributions in determining the QCD critical point of relativistic heavy-ion collisions, it is essential to take the dynamical evolution.

\end{abstract}


\maketitle

\section{Introduction}
\label{sec:intro}
In recent years, the study on the structure of the QCD phase diagram has been a key research area of relativistic heavy-ion collisions~\cite{Adams:2005dq,Bzdak:2019pkr}. Lattice QCD calculations indicate that the phase transition from quark gluon plasma to hadron gas is  smooth when the baryon chemical potential $\mu_{B}$ disappears~\cite{Aoki:2006we,deForcrand:2002hgr,Ding:2015ona}; however, this transition is a first order phase transition at a large $\mu_{B}$ ~\cite{Ejiri:2008xt}. As an analog of the critical point's role in the phase diagram of water, the critical point, which is located at the end point of the first order phase transition line connected with the crossover region, is a significantly important feature of the QCD phase diagram~\cite{Stephanov:1998dy,Stephanov:1999zu,Stephanov:2004wx,Asakawa:1989bq,Barducci:1989wi,Berges:1998rc,Halasz:1998qr,Berdnikov:1999ph,Fodor:2001pe,Hatta:2002sj}.  Several ideas have been proposed to search for the QCD critical point, which include approaches via ratio of HBT-radii~\cite{Lacey:2014wqa}, yield ratio of light nuclei~\cite{Sun:2017xrx,Yu:2018kvh,Shao:2019xpj}, and directed flow slope of net proton~\cite{Stoecker:2004qu}. It has been suggested that the moments of conserved charges are proportional to some powers of the correlation length closely related to  susceptibilities~\cite{Stephanov:2008qz,Athanasiou:2010kw,Stephanov:2011pb,Gavai:2010zn,Cheng:2008zh}. Because susceptibilities diverge at the critical point and the fluctuations of conserved charges are proportional to their corresponding susceptibilities, moments and moment products of conserved charges have been proposed as sensitive probes to search for the QCD critical point~\cite{Stephanov:2008qz,Luo:2011ts,Luo:2011rg,Luo:2010by}.  

In Lattice QCD, the susceptibilities of conserved charges, e.g. baryon number $B$, strangeness $S$ and electric charge $Q$ , are defined as
\begin{equation}
\chi_{lmn}^{BSQ}=\frac{\partial^{\,l+m+n}(P/T^4)}{\partial(\mu_{B}/T)^{l}\partial(\mu_{S}/T)^{m}\partial(\mu_{Q}/T)^{n}},
\end{equation}
where P is pressure, T is temperature, $\mu_{BSQ}$ are chemical potentials of conserved charges (baryon number $B$, strangeness $S$, and electric charge $Q$). To connect the susceptibilities with the measured products of moments of the distributions of the corresponding conserved charges, we can conveniently introduce the following volume-independent ratios:
\begin{equation}
\chi_2/\chi_1=\sigma^2/M;
\chi_3/\chi_{2}=~S\sigma;
\chi_4/\chi_{2}=\kappa\sigma^2.
\label{moments}
\end{equation}

Here, the moments of mean($M$), standard deviation($\sigma$), skewness($S$), and kurtosis($\kappa$) are measurable experimental observerbles, defined as follows~\cite{Luo:2011ts,Adamczyk:2017wsl,Luo:2013bmi,Sahoo:2012wn}:  
\begin{equation}\label{mean}
M=K_1=\langle X\rangle,
\end{equation}
\begin{equation}\label{sigma}
\sigma=\sqrt{K_2}=\sqrt{\langle (X-M)^{2}\rangle },
\end{equation}
\begin{equation}\label{S}
S=\frac{K_3}{K_2^{3/2}}=\frac{\langle(X-M)^{3}\rangle} {\sigma^3},
\end{equation}
\begin{equation}\label{K}
\kappa=\frac{K_4}{K_2^{2}}=\frac{\langle (X-M)^{4}\rangle} {\sigma^4}-3,
\end{equation}
where $X$ is the net charge, i.e. the difference between the positive and negative conserved charges, $K_n$ is the $n$th cumulant of the net-charge distribution, e.g. $K_n=\langle(X-M)^{n}\rangle$ for n=1,2 and 3, and $K_4=\langle(X-M)^{4}\rangle-3\langle(X-M)^{2}\rangle^{2}$, and $\langle ... \rangle$ means the average value obtained over all events. 

The BNL Relativistic Heavy Ion Collider (RHIC) has been performing a Beam Energy Scan program~\cite{Aggarwal:2010wy, Aggarwal:2010cw, Adamczyk:2017iwn} with the task to search for the critical point of the QCD phase diagram~\cite{Stephanov:1998dy}. Several moment types of net-proton, net-strangeness and net-charge have been reported~\cite{Aggarwal:2010wy,Adamczyk:2014fia,Adamczyk:2017wsl,Adamczyk:2013dal,Adam:2021fby,Pandav:2020uzx,Abdallah:2021fzj}. As the proxy of baryon number, a non-monotonic energy dependence $\kappa\sigma$ of the net-proton multiplicity distribution has been observed from the first phase of BES measurements, which is consistent with the expected QCD critical fluctuations~\cite{Adam:2021fby,Abdallah:2021fzj}. However, for the proxy of strangeness, the measured net-kaon cumulant ratios do not exhibit any significant non-monotonic energy dependence~\cite{Adamczyk:2017wsl}. In 2014, the STAR collaboration published the first measurement on the moments of the net-charge ($X \equiv Q_{+}-Q_{-}$) distribution fluctuation~\cite{Adamczyk:2014fia}, which deviated from the expectations of Poisson and negative binomial distributions; however no non-monotonic behavior was observed as a function of the colliding energy. 

In this study, we focus on the dynamical evolution of the moments of net-charge multiplicity distribution using a multi-phase transport model, because heavy-ion collisions consist of dynamical evolution stages that exist between the Lattice QCD calculation and final experimental measurement. This paper is organized as follows. We introduce the AMPT model and our method of calculating moments in Section II. Our results and discussions are presented in Section III. Finally, a summary is presented in Section IV.

\section{Model and calculation method}
\subsection{The AMPT model}
\label{sec:partA}
A multiphase transport model (AMPT), has been widely used to investigate several aspects of the physics of relativistic heavy-ion collisions~\cite{Lin:2004en,Ma:2016fve,Ma:2013gga,Ma:2013uqa,Bzdak:2014dia}. The AMPT model with a string melting mechanism~\cite{Lin:2004en} primarily comprises four stages: initial condition, partonic cascade, hadronization and hadronic rescatterings. The initial condition obtained from the HIJING model~\cite{Wang:1991hta,Gyulassy:1994ew} primarily provides the spatial and momentum distributions of minijet partons and soft strings. Under the string melting mechanism, both excited strings and minijet partons are melted into partons, i.e. decomposed into constituent quarks according to their flavor and spin structures.  The strong interactions among partons are simulated by Zhang's parton cascade model~\cite{Zhang:1997ej}, which includes elastic partonic scatterings with a fixed cross section (3 mb). A simple quark coalescence model describes the conversion of these partons to hadrons, i.e. hadronization. The ART model simulates interactions among the hadrons and corresponding inverse reactions, as well as resonance decays~\cite{Li:1995pra}. In old versions of AMPT model, charge conservation is violated due to certain problems~\cite{Lin:2014uwa}. In this study, a new version of the AMPT model which ensures the strict charge conservation of electric charge for each hadronic reaction channel, is used to investigate the net-charge fluctuations. In order to study the energy dependence of the moments of net-charge fluctuations, we simulated Au+Au minus bias collisions for seven different energies ($\sqrt{s_{NN}}$=200, 62.4, 39, 27, 19.6, 11.5, 7.7 GeV) of the BES program at RHIC. 

\subsection{Calculation method}
\label{sec:partB}
The mean($M$), standard deviation($\sigma$), skewness($S$), and kurtosis($\kappa$) can be used to characterize different features of the multiplicity distribution. The mean represents the average value of the distribution, standard deviation represents the degree of dispersion of the distribution, the skewness represents the asymmetry of the distribution, and the kurtosis describes the degree to which the distribution is peaked relative to the normal distribution. The formulas of $M$, $\sigma$, $S$ and $\kappa$ are shown in Eqs.~(\ref{mean}) -(\ref{K}). To calculate the aforementioned different moments of net-charge multiplicity distribution, we select charged particles with transverse momenta $0.2 <p_{T}< 2.0$ GeV/c and a pseudorapidity range of $-0.5<\eta<0.5$. We adopt the charged particle multiplicity distribution to define centrality bins. To avoid the self-correlation effect(ACE)~\cite{Luo:2013bmi,Chatterjee:2019fey}, we select the $\eta$ range of charged particles, which are used to define centrality bins to exclude $|\eta|<0.5$. We apply the Delta method to estimate the statistical errors of moments or moment products, according to  the procedure of the experimentalists in~\cite{Luo:2011tp}.

In addition, the initial volume fluctuations can contribute to the moments as a background, which can induce a centrality bin width effect(CBWE)~\cite{Luo:2011ts,Luo:2013bmi} and a centrality resolution effect(CRE)~\cite{Luo:2013bmi}. The CBWE is triggered by the finite width of centrality bin. In other words, if the width of centrality bin is sufficiently small , the influence of the CBWE will be weakened. The CRE originates from the uncertainty of the collision centrality determination, as different initial collision geometries will trigger the uncertainty of results. To reduce the contribution of these background effects, the centrality bin width correction (CBWC) and the centrality resolution correction (CRC) are applied in a similar manner to the experiments in~\cite{Luo:2011ts,Luo:2013bmi,Sahoo:2012wn}. We compare the results with and without the CBWC and CRC in subsection ~\ref{sec:partC} and ~\ref{sec:partD}, respectively.

\section{Results and Discussions}
\label{sec:results}

\subsection{Centrality bin width effect and correction}
\label{sec:partC}

To eliminate the influence of the CBWE, CBWC has been proposed~\cite{Luo:2011ts,Luo:2013bmi,Sahoo:2012wn}. This method is defined as:

\begin{equation}\label{X}
X=\frac{\sum_{r}n_{r}X_{r}} {\sum_{r}\ n_{r}},
\end{equation}
where $X$ represents any moments or moment products, such as $M$, $\sigma$, and $S\sigma$, $n_{r}$ is the number of events in the $r${th} multiplicity bin, and $X_{r}$ represents moments or moment products in the $r${th} multiplicity bin.

\begin{figure}
\centering
\includegraphics[scale=0.45]{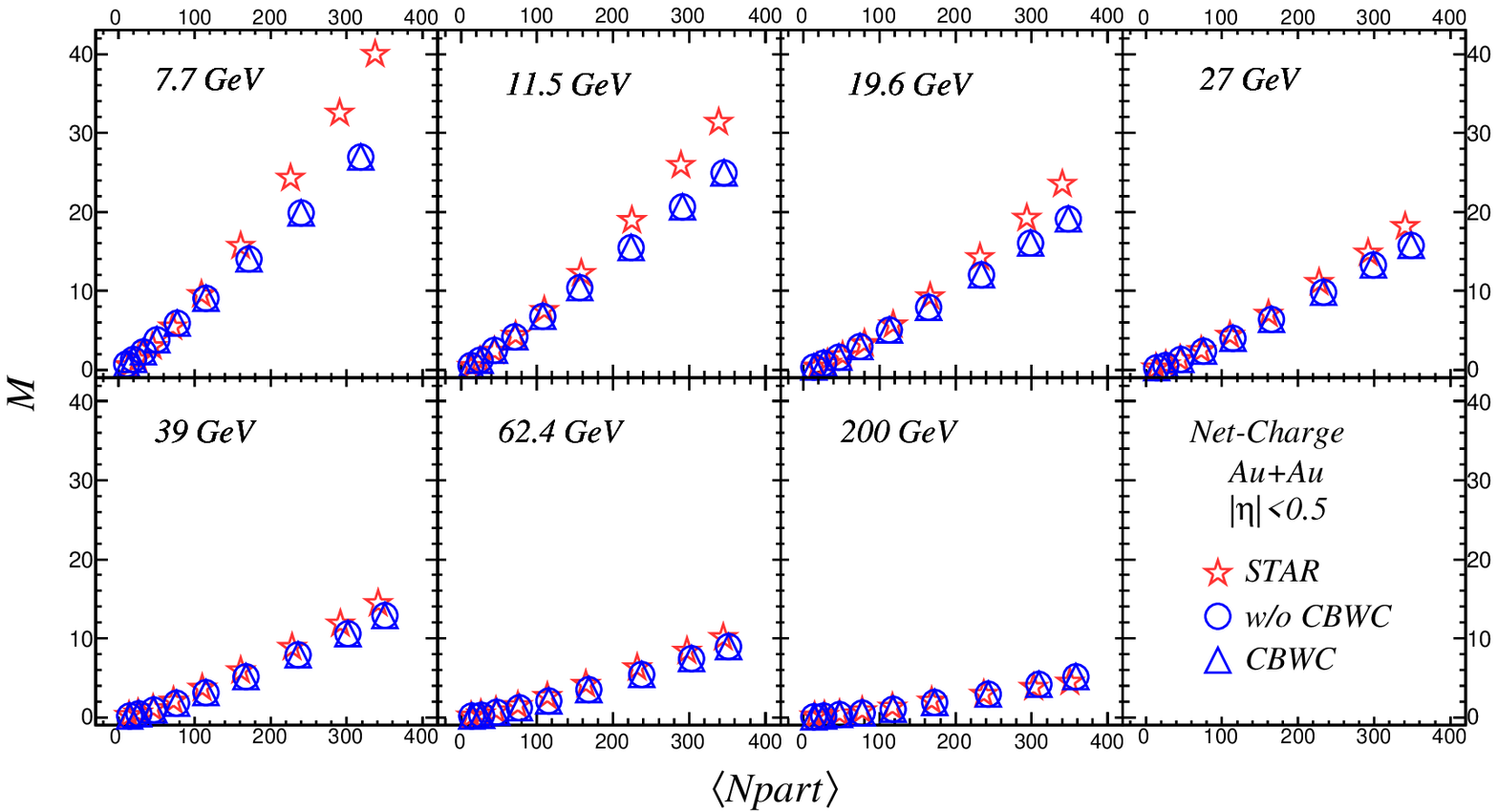}
\includegraphics[scale=0.45]{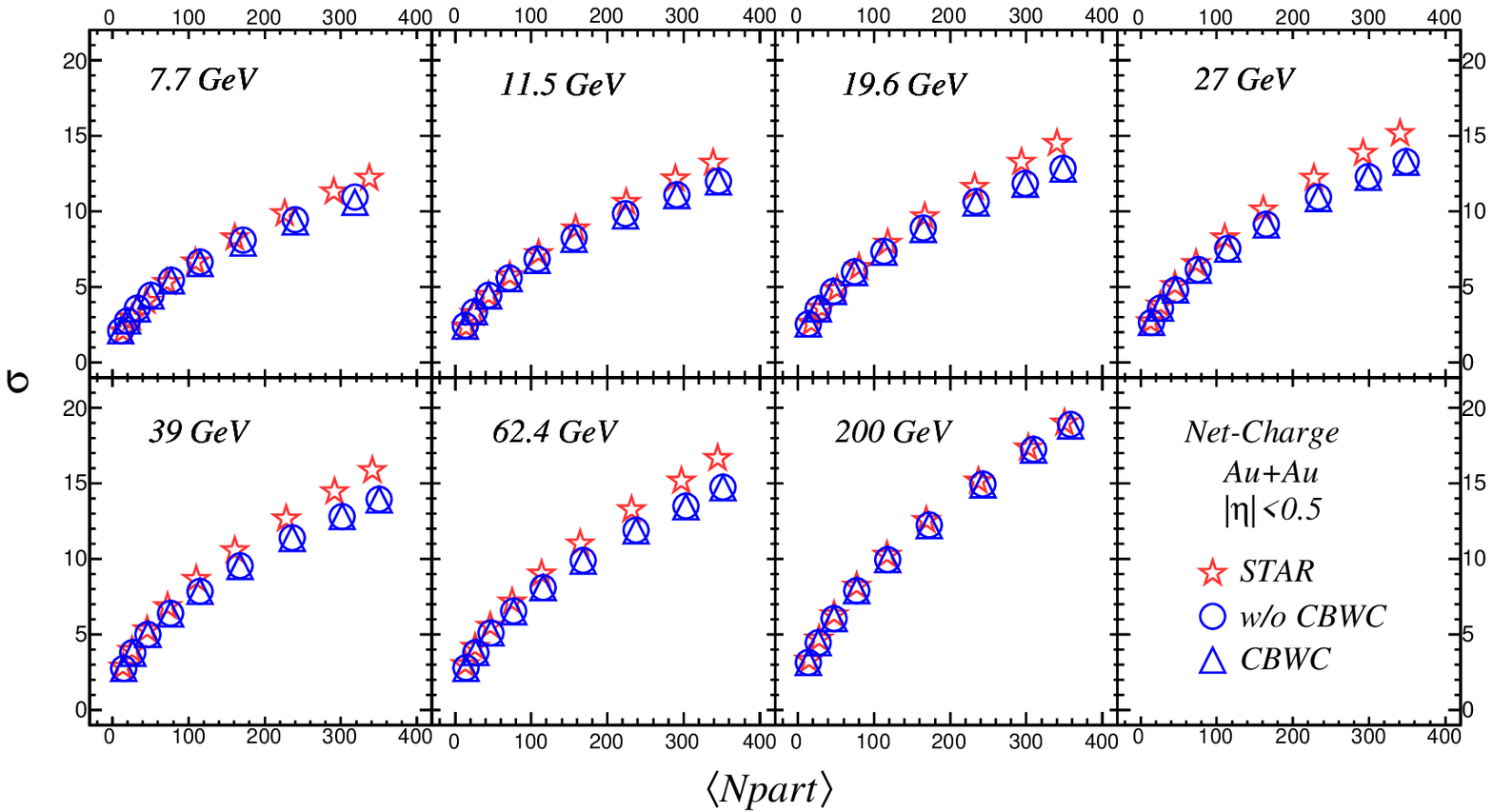}
\caption{(Color online) Centrality dependence of mean($M$)(top plots) and standard deviation($\sigma$)(bottom plots) for net-charge multiplicity distributions in Au+Au collisions at $\sqrt{s_{NN}}=7.7-200\,\mathrm{GeV}$, where the STAR data are obtained from Ref.~\cite{Adamczyk:2014fia}.
}
\label{fig-1}
\end{figure}

In order to clearly see the influence of the CBWE on moments, the results without and with the CBWC are compared. Figure~\ref{fig-1} presents the centrality dependences of mean($M$) and standard deviation($\sigma$) of the net-charge multiplicity distributions in Au+Au collisions at seven different colliding energies($\sqrt{s_{NN}}=7.7-200\,\mathrm{GeV}$), which are presented as a function of the average number of participant nucleons $\langle N_{part}\rangle$. The moments $M$ and $\sigma$ without the CBWC are consistent with those with the CBWC, which indicates that the CBWE has negligible influence on the mean and standard deviation of the net-charge distribution. As the $\langle N_{part}\rangle$ increases, the $M$ and $\sigma$ both exhibit an increasing dependence of $\langle N_{part}\rangle$ for each energy. When the colliding energy increases, the mean decreases but the standard deviation increases for a given $\langle N_{part}\rangle$ value. This indicates that the QCD matter created at higher energies is in a state with less electric charge chemical potential but with larger fluctuation. 

\begin{figure}
\centering
\includegraphics[scale=0.45]{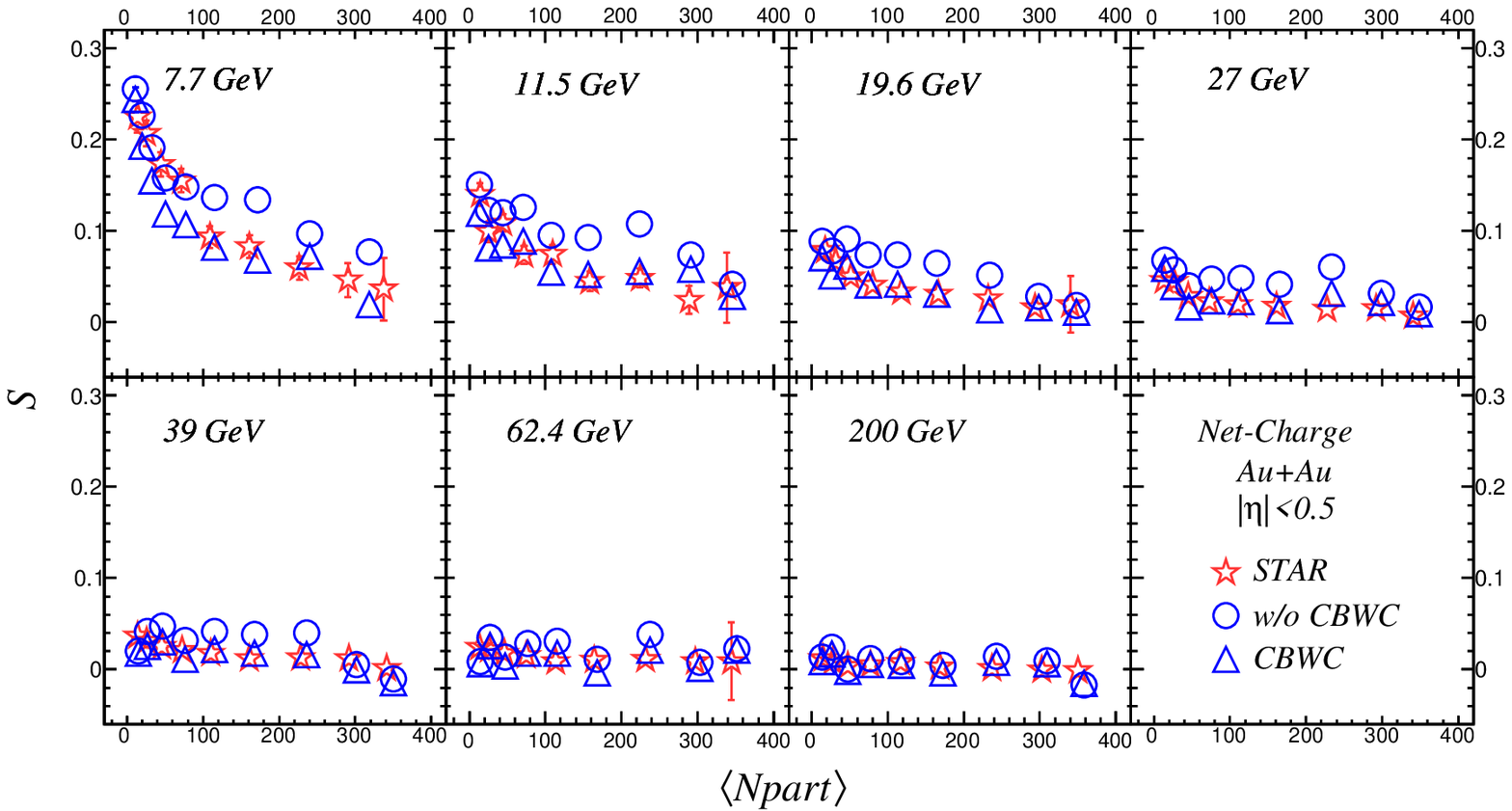}
\includegraphics[scale=0.45]{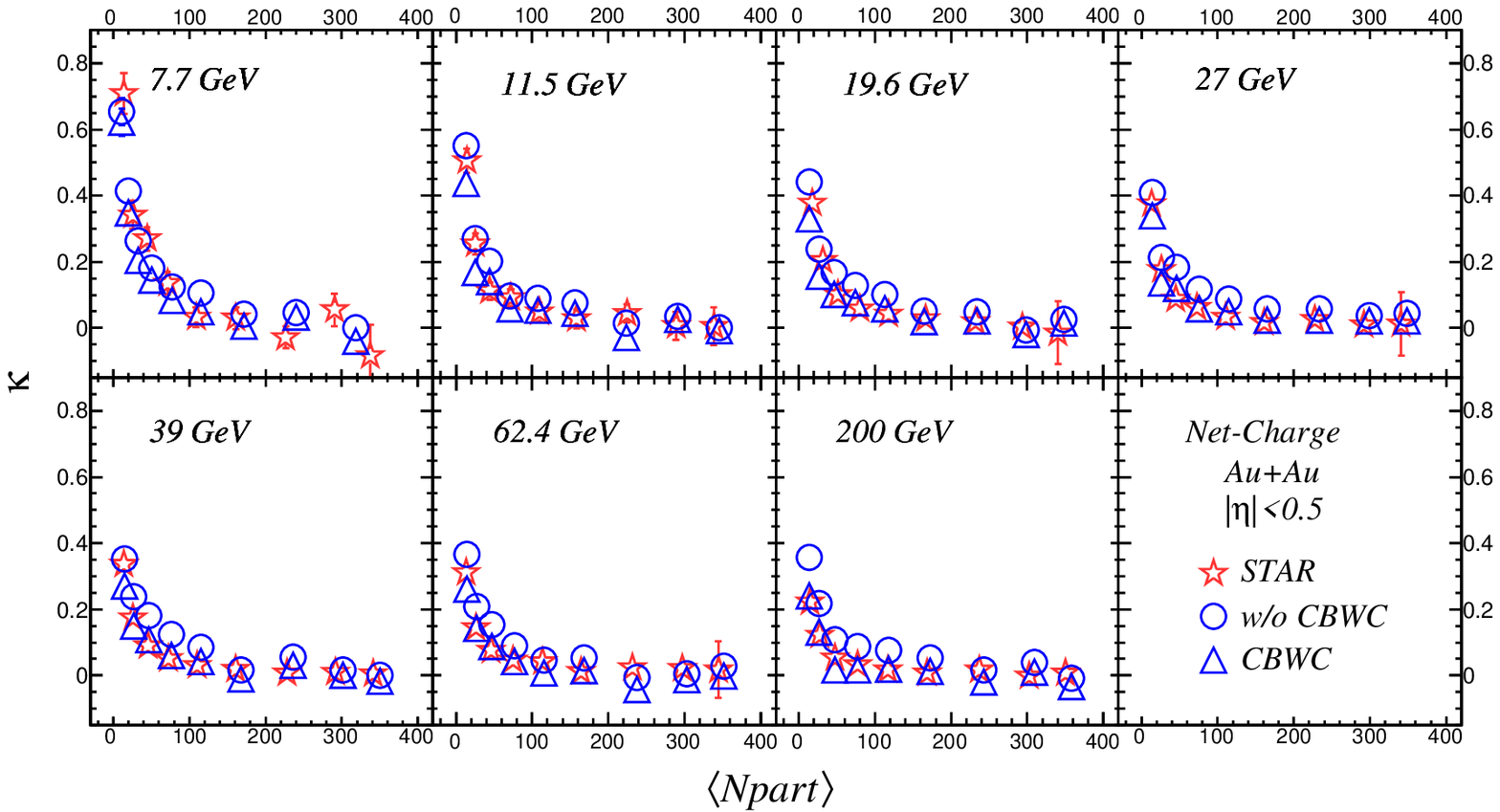}
\caption{(Color online)  Centrality dependence of skewness($S$)(top plots) and kurtosis($\kappa$)(bottom plots) for net-charge multiplicity distributions in Au+Au collisions at $\sqrt{s_{NN}}=7.7-200\,\mathrm{GeV}$, where the STAR data are obtained from Ref.~\cite{Adamczyk:2014fia}
}
\label{fig-2}
\end{figure}

Figure~\ref{fig-2} illustrates the centrality dependences of the skewness($S$) and kurtosis($\kappa$) for net-charge multiplicity distributions in Au+Au collisions at $\sqrt{s_{NN}}=7.7-200\,\mathrm{GeV}$. For both results of skewness and kurtosis, the values with the CBWC are systematically smaller than those without the CBWC. However, when the energy increases, the difference between the results with and without the CBWC becomes  increasingly smaller. These results with the CBWC can describe experimental data better than those without the CBWC, which indicates that the CBWE makes a significant contribution to skewness and kurtosis. The CBWE has a larger influence on the low energies than on the high energies, which indicates that the effect of volume fluctuations inside one centrality bin increases with the decreasing of colliding energy. Therefore, the CBWC is very essential, especially for low energies. 

\begin{figure}
\centering
\includegraphics[scale=0.45]{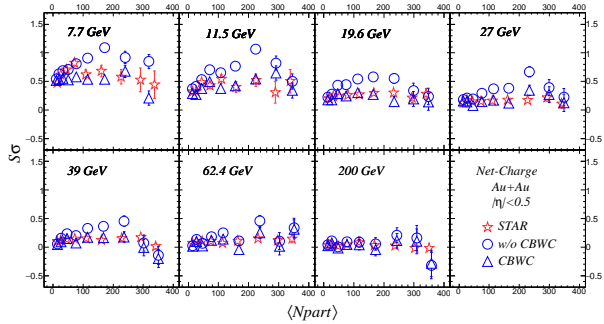}
\caption{(Color online)  Centrality dependence of moment products $S\sigma$ for net-charge multiplicity distributions in Au+Au collisions at $\sqrt{s_{NN}}=7.7-200\,\mathrm{GeV}$, where the STAR data are obtained from Ref.~\cite{Adamczyk:2014fia}.
}
\label{fig-3}
\end{figure}

Figure~\ref{fig-3} presents the centrality dependence of $S\sigma$ on net-charge multiplicity distributions in Au+Au collisions at $\sqrt{s_{NN}}=7.7-200\,\mathrm{GeV}$. It can be observed that the moment product $S\sigma$ of net charge distribution with the CBWC can describe experimental data better than the $S\sigma$ without the CBWC which overestimates the experimental data. This can be easily understood because $S$ with the CBWC does a better job for describing the experimental data than without the CBWC, as shown in Fig.~\ref{fig-2}.

\begin{figure}
\centering
\includegraphics[scale=0.45]{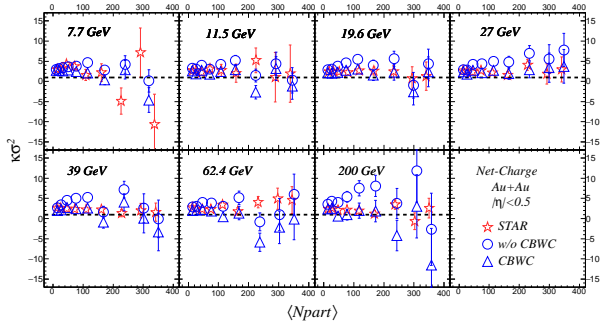}
\caption{(Color online)  Centrality dependence of moment products $\kappa\sigma^{2}$ for net-charge multiplicity distributions in Au+Au collisions at $\sqrt{s_{NN}}=7.7-200\,\mathrm{GeV}$, where the STAR data are obtained from Ref.~\cite{Adamczyk:2014fia}.
}
\label{fig-4}
\end{figure}

Figure~\ref{fig-4} presents the dependence results of the $\kappa\sigma^{2}$ on net-charge multiplicity distributions in Au+Au collisions at $\sqrt{s_{NN}}=7.7-200\,\mathrm{GeV}$. The AMPT results with and without the CBWC can both describe the experimental data within the statistical errors. However, the $\kappa\sigma^{2}$ with the CBWC is slightly smaller than that without the CBWC, which is approximately the constant of unity within the statistical errors. 

\subsection{The centrality resolution effect and correction}
\label{sec:partD}

We also study the influence of the centrality resolution effect (CRE) on the moments and moment products of net-charge multiplicity distributions in Au+Au collisions. Because the CRE is induced by initial volume fluctuations of impact parameter or collision centrality, using different $\eta$ ranges to define collision centrality is considered a possible way to weaken this effect. It should be stated that to avoid the possible interference of the CBWE, all the AMPT results presented in this subsection are obtained with the CBWC. 

\begin{figure}
\centering
\includegraphics[scale=0.45]{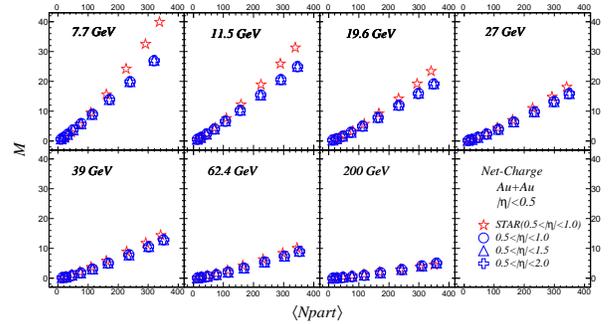}
\includegraphics[scale=0.45]{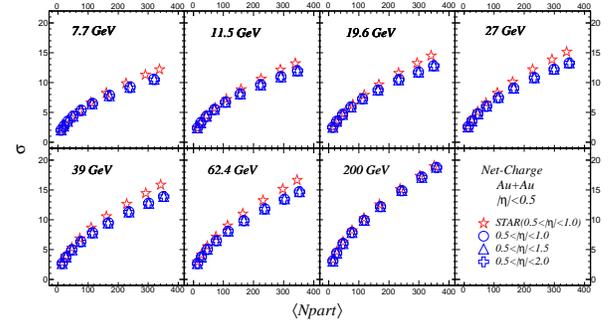}
\caption{(Color online)  Centrality dependence of mean($M$)(top plots) and standard deviation($\sigma$)(bottom plots) for net-charge multiplicity distributions in Au+Au collisions at $\sqrt{s_{NN}}=7.7-200\,\mathrm{GeV}$ with different centrality definitions, where the STAR data are obtained from Ref.~\cite{Adamczyk:2014fia}.
}
\label{fig-5}
\end{figure}

Figure~\ref{fig-5} illustrates that the centrality dependence of the mean($M$) and standard deviation($\sigma$) for net-charge multiplicity distributions with different centrality definitions in Au+Au collisions at $\sqrt{s_{NN}}=7.7-200\,\mathrm{GeV}$. To define centrality bins, the charge multiplicities from three different pseudorapidity ranges (0.5$<|\eta|<$1.0, 0.5$<|\eta|<$1.5 and 0.5$<|\eta|<$2.0) are adopted, all of which are out of midpseudorapidity such that possible autocorrelations are avoided. We find that the AMPT results of mean and standard deviation with three different centrality definitions are overlapped and approximate to the experiment data. This indicates that the CRE has little influence on the $M$ and $\sigma$ of net-charge multiplicity distributions from the AMPT model. 

\begin{figure}
\centering
\includegraphics[scale=0.45]{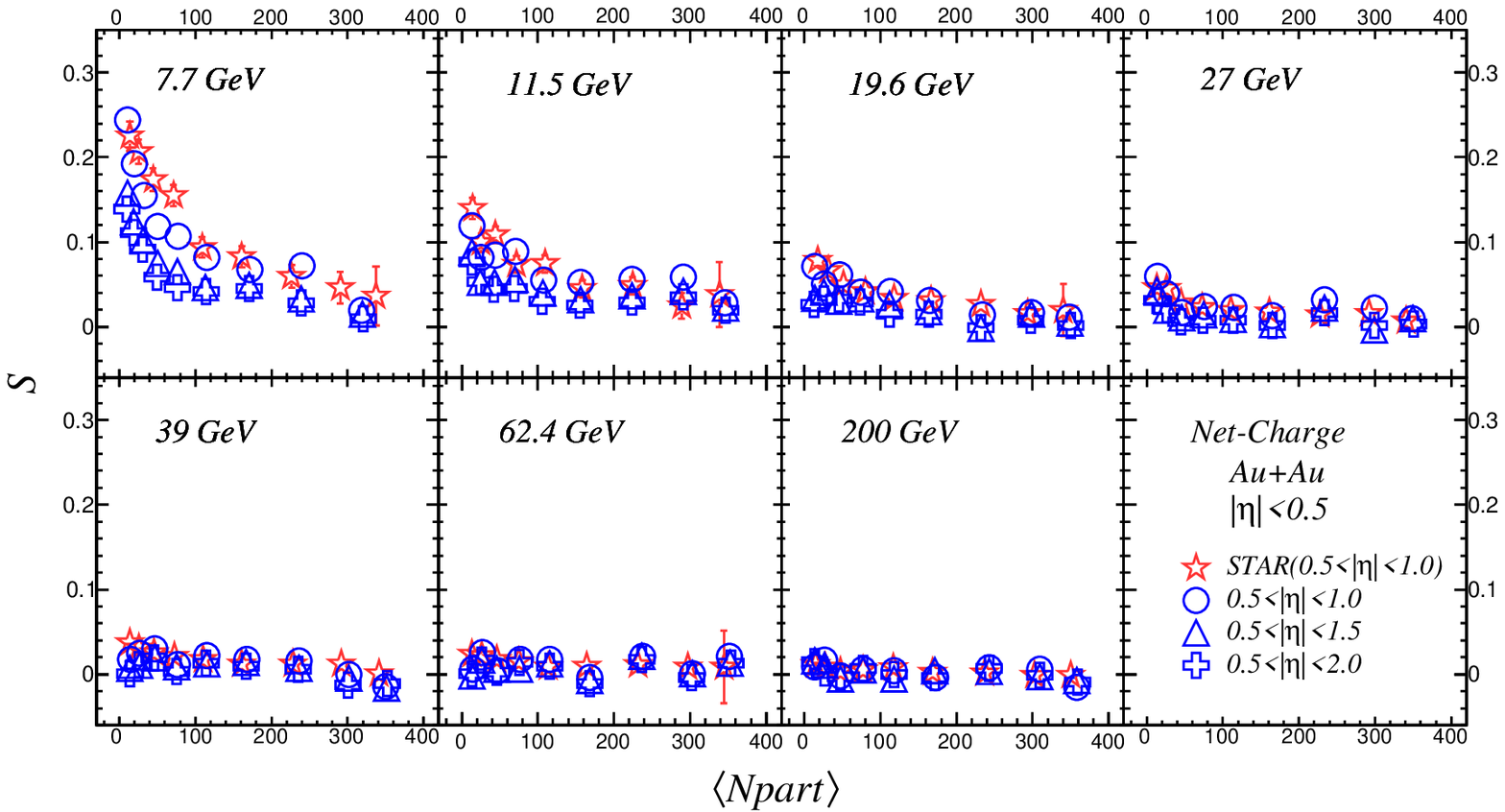}
\includegraphics[scale=0.45]{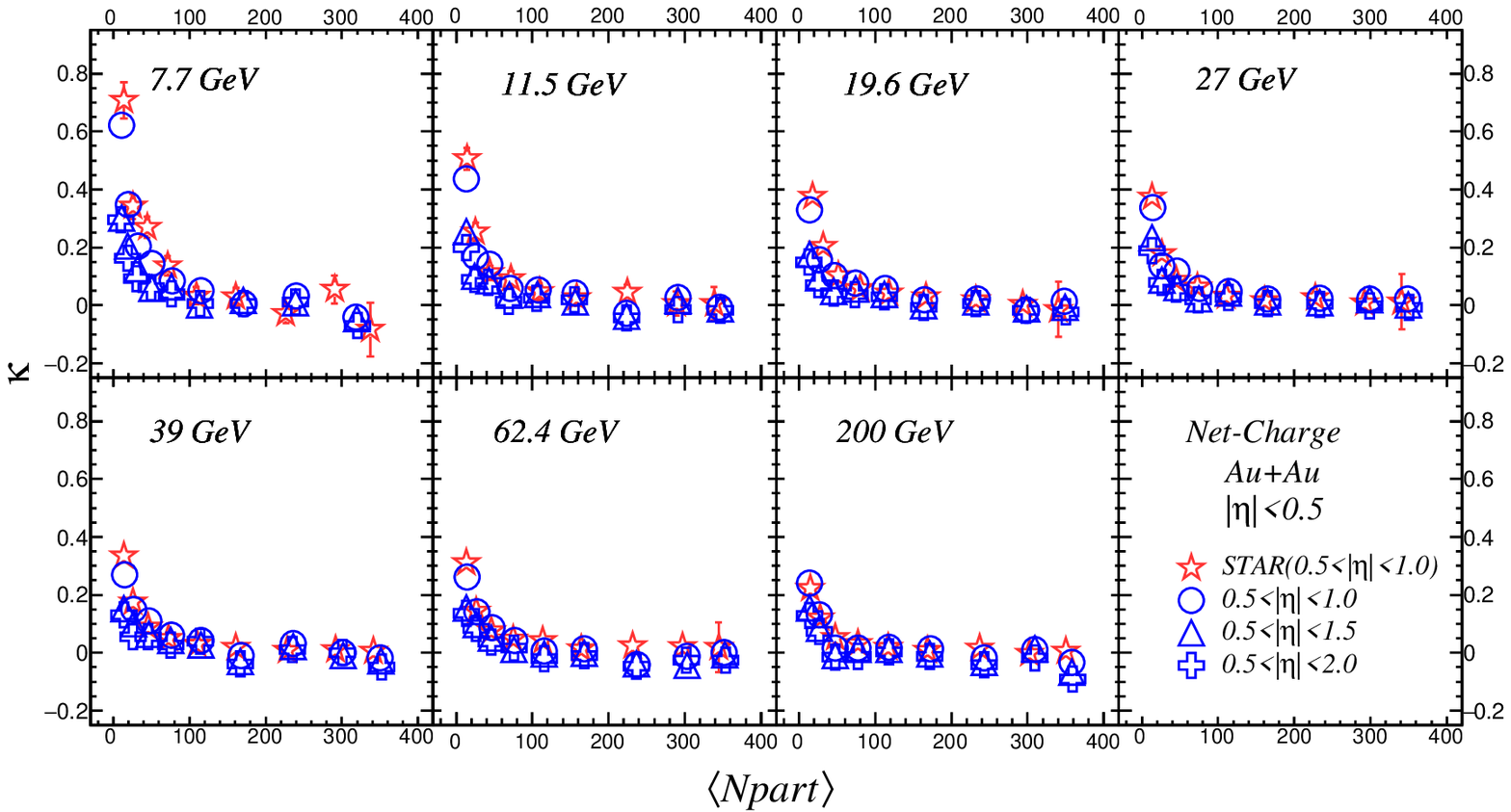}
\caption{(Color online)  Centrality dependence of skewness($S$)(top plots) and kurtosis($\kappa$)(bottom plots) for net-charge multiplicity distributions in Au+Au collisions at $\sqrt{s_{NN}}=7.7-200\,\mathrm{GeV}$ with different centrality definitions, where the STAR data are obtained from Ref.~\cite{Adamczyk:2014fia}.
}
\label{fig-6}
\end{figure}

Figure~\ref{fig-6} presents that the centrality dependence of skewness($S$) and kurtosis($\kappa$) for the net-charge multiplicity distributions with different centrality definitions in Au+Au collisions at $\sqrt{s_{NN}}=7.7-200\,\mathrm{GeV}$. Regarding skewness, the AMPT results can describe the experimental data well when we select the centrality definition as same as in experiment. However, when the $\eta$ range increases below 27 GeV, we observe that the value of $S$ decreases and seems to saturate for $0.5<|\eta|<1.5$, which indicates the centrality resolution effect will enhance the skewness value of net-charge multiplicity distributions. From 27 GeV to 200 GeV, the differences between the results from three different centrality definitions appear increasingly smaller. This indicates that the CRE has a larger influence on the value of skewness for lower energies. For kurtosis, the AMPT results from different centrality definitions are all similar to the experimental data. This indicates that the CRE has no significant contribution to the value of kurtosis. 

\begin{figure}
\centering
\includegraphics[scale=0.45]{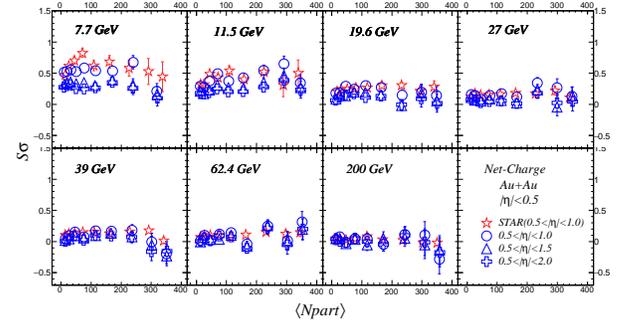}
\caption{(Color online)  Centrality dependence of moment products $S\sigma$ for net-charge multiplicity distributions in Au+Au collisions at $\sqrt{s_{NN}}=7.7-200\,\mathrm{GeV}$ with different centrality definitions, where the STAR data are obtained from Ref.~\cite{Adamczyk:2014fia}.
}
\label{fig-7}
\end{figure}

Figure~\ref{fig-7} presents that the centrality dependence of moment products($S\sigma$) for the net-charge multiplicity distributions with different centrality definitions in Au+Au collisions at $\sqrt{s_{NN}}=7.7-200\,\mathrm{GeV}$. At low energies(below 27 GeV), as the $\eta$ range increases, the value of $S\sigma$ decreases, which can be understood as a result of the CRE on $S$. The $S\sigma$ describes the experimental data well when the $\eta$ range of the centrality definition is $0.5<|\eta|<1.0$. At high energies(above 27 GeV), the values of $S\sigma$ from three different centrality definitions all appear similar to experimental data.  

\begin{figure}
\centering
\includegraphics[scale=0.45]{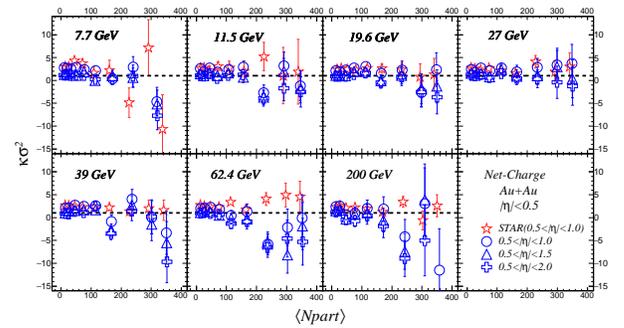}
\caption{(Color online)  Centrality dependence of moment products $\kappa\sigma^{2}$ for net-charge multiplicity distributions in Au+Au collisions at $\sqrt{s_{NN}}=7.7-200\,\mathrm{GeV}$ with different centrality definitions, where the STAR data are obtained from Ref.~\cite{Adamczyk:2014fia}.
}
\label{fig-8}
\end{figure}

Figure~\ref{fig-8} presents that the centrality dependence of $\kappa\sigma^{2}$ for the net-charge multiplicity distributions with different centrality definitions in Au+Au collisions at $\sqrt{s_{NN}}=7.7-200\,\mathrm{GeV}$. From low to high energy,  the values of $\kappa\sigma^{2}$ from  three different centrality definitions are all approximately the constant of unity, which are all consistent with the experimental data within errors.

\subsection{Energy dependence}

\begin{figure}
\centering
\includegraphics[scale=0.45]{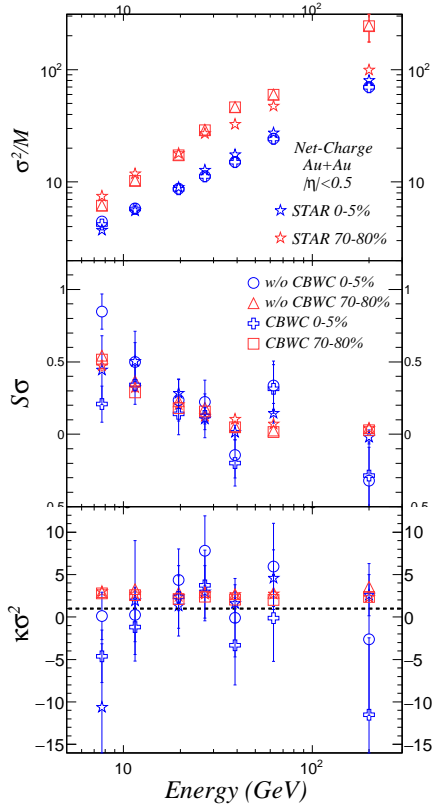}
\caption{(Color online)  Beam-energy dependence of moment products $\sigma^{2}/M$, $S\sigma$, and $\kappa\sigma^{2}$ in most central ($0-5\%$) and peripheral ($70-80\%$) bins of Au+Au collisions, where the STAR data are obtained from Ref.~\cite{Adamczyk:2014fia}. 
}
\label{fig-9}
\end{figure}

\begin{figure}
\centering
\includegraphics[scale=0.45]{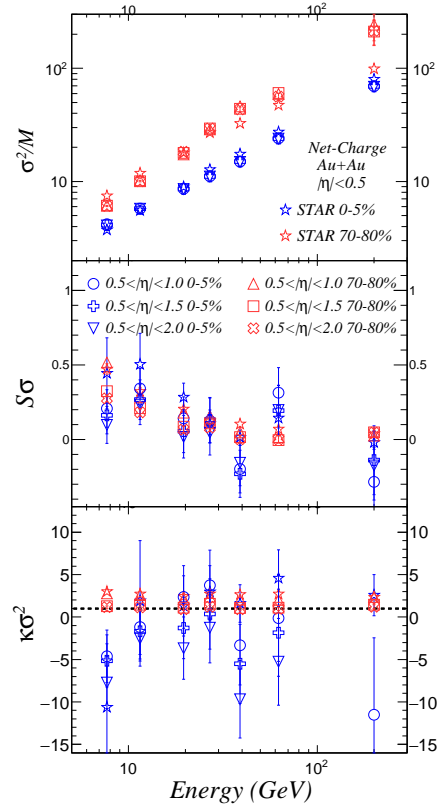}
\caption{(Color online)  Beam-energy dependence of moment products $\sigma^{2}/M$, $S\sigma$, and  in most central ($0-5\%$) and peripheral ($70-80\%$) bins of Au+Au collisions with different centrality definitions, where the STAR data are obtained from Ref.~\cite{Adamczyk:2014fia}. 
}
\label{fig-10}
\end{figure}

As mentioned above, it is believed that moment products, which can eliminate the volume fluctuation effect, are sensitive to the critical fluctuations. Hence, we study the energy dependence of moment products, as presented in Figs.~\ref{fig-9} and ~\ref{fig-10}, where both AMPT results with and without the CBWC, are presented in comparison with the experimental data.

Figure~\ref{fig-9} illustrates the energy dependence of the moment products, which includes $\sigma^{2}/M$, $S\sigma$, and $\kappa\sigma^{2}$. Regarding $\sigma^{2}/M$, this moment product exhibits an exponential response to the energy for a given collision centrality. The $\sigma^{2}/M$ for peripheral centrality bin ($70-80\%$) is greater than that for central centrality bin ($0-5\%$). In addition, the values of $\sigma^{2}/M$ are similar to the experimental data whether the CBWC is considered or not. For $S\sigma$ and $\kappa\sigma^{2}$, we observe that $S\sigma$ decreases as energy increases; however, $\kappa\sigma^{2}$ is independent of energy. The values of $S\sigma$ or $\kappa\sigma^{2}$ with the CBWC appear systematically smaller than those without the CBWC. This is consistent with the above result, which indicates that although the CBWE has little influence on $\sigma$, it enhances $S$ and $\kappa$ slightly.

To study the influence of the CRE on the energy dependence of the moment products, different centrality definitions have been applied. Figure~\ref{fig-10} presents the beam-energy dependence of moment products $\sigma^{2}/M$, $S\sigma$, and $\kappa\sigma^{2}$ with the CBWC for most central ($0-5\%$) and peripheral ($70-80\%$) centrality bins with three different centrality definitions. For $\sigma^{2}/M$, the results with different centrality definitions are almost same, which indicates the CRE has little influence on $\sigma^{2}/M$. However, for $S\sigma$ and $\kappa\sigma^{2}$, the value of the two moment products systematically become smaller as the $\eta$ range becomes wider, which is consistent with the above result, which indicates that although the CRE has little influence on $M$ and $\sigma$, it enhances $S$ and $\kappa$ slightly.

In general, after considering the CBWC and CRE, the AMPT results of moment products of net-charge multiplicity distribution are consistent with the experimental measurements. No non-monotonic energy dependence is observed, which is not surprising because no QCD critical fluctuations are included in the AMPT model at all. 

\subsection{Stage evolution}
It is well known that heavy-ion collisions are actually a complicated dynamical evolution comprising several important evolution stages. Therefore, it is important to study the stage evolution of moment products, which can help us understand the dynamics of the fluctuation observables.  This complements to the results from the Lattice QCD which basically belongs to a static thermal approach without evolution dynamics. Using the AMPT model, we focus on the moment products at four evolution stages (i.e. initial stage, after parton cascade, after coalescence, and after hadron rescatterings) in Au+Au collisions for two typical energies, 7.7 GeV and 200 GeV.

To elucidate the possible contribution from any dynamics to the moments or moment products, a reference of Poisson baseline is crucially required. To obtain the expected moments of net-charge distribution from two independent Poisson distributions of positively and negatively charged particles, we first calculate $K_n^{Q_+}$ and $K_n^{Q_-}$ for the multiplicity distributions of positively and negatively charged particles ($Q_+$ and $Q_-$), then obtain the Poisson expectation of net charge $K_n^{Poisson}$=$K_n^{Q_+}+(-1)^n K_n^{Q_-}$, which reflects the case without any correlation between positively and negatively charged particles~\cite{Luo:2017faz,Tarnowsky:2012vu}. To illustrate the close relation between fluctuation and correlation, $\Delta K_n=K_2^{Poisson}-K_2^{netq}$ is defined to disclose the difference between the net-change cumulant $K_n^{netq}$ and their corresponding Poisson expectation $K_n^{Poisson}$, with the following relations with correlation functions:
\begin{eqnarray}
\label{C2netq}
\Delta K_2 &\simeq&2C_{2}^{(1,1)} 
\notag \\
&\simeq&2 (\langle Q_- Q_+ \rangle-  \langle Q_- \rangle \langle Q_+ \rangle),
\end{eqnarray}

\begin{eqnarray}
\label{C3netq}
\Delta K_3 &\simeq&3C_{3}^{(2,1)}-3C_{3}^{(1,2)}
\notag \\
&\simeq&3 (\langle Q_- Q_+^2\rangle- \langle Q_-\rangle\langle Q_+^2\rangle)
\notag \\
&&-3 ( \langle Q_-^2 Q_+\rangle- \langle Q_-^2\rangle \langle Q_+\rangle)
\notag \\
&&+6 \langle Q_-\rangle(\langle Q_- Q_+\rangle- \langle Q_-\rangle \langle Q_+\rangle)
\notag \\
&&- 6\langle Q_+\rangle(\langle Q_- Q_+\rangle- \langle Q_-\rangle\langle Q_+\rangle) ,
\end{eqnarray}

\begin{eqnarray}
\label{C4netq}
\Delta K_4 &\simeq&2C_{2}^{(1,1)}+6C_{3}^{(2,1)}+6C_{3}^{(1,2)}
\notag \\
&&+4C_{4}^{(3,1)}+4C_{4}^{(1,3)}-6C_{4}^{(2,2)} ,
\end{eqnarray}

where where $C_{n+m}^{(n,m)}$ represents $n+m$ correlation function for $n$ positively charged particles and $m$ negatively charged particles, which can be calculated via the relations with factorial moments (See Ref.~\cite{Bzdak:2016sxg} for details). Note that we only focus on the correlation between positively and negatively charged particles and ignore the correlation between same-sign charged particles. Even with the assumption, we could not explicitly express the formula for Eq.~(\ref{C4netq}) because it is very complicated. It can be easily observed that the differences between the net-charge moments and their Poisson expectations originate from two-particle, three-particle or four-particle correlations between positively and negatively charged particles.

%%%%%%%%%%%%%%%%%%%%%%%%%  $S\sigma$  %%%%%%%%%%%%%%%%%%%%%%%%%%%%%%%%%%%%%%%%%%%%
\begin{figure*}[htb]
\begin{minipage}[t]{80mm}
 \includegraphics[width=80mm]{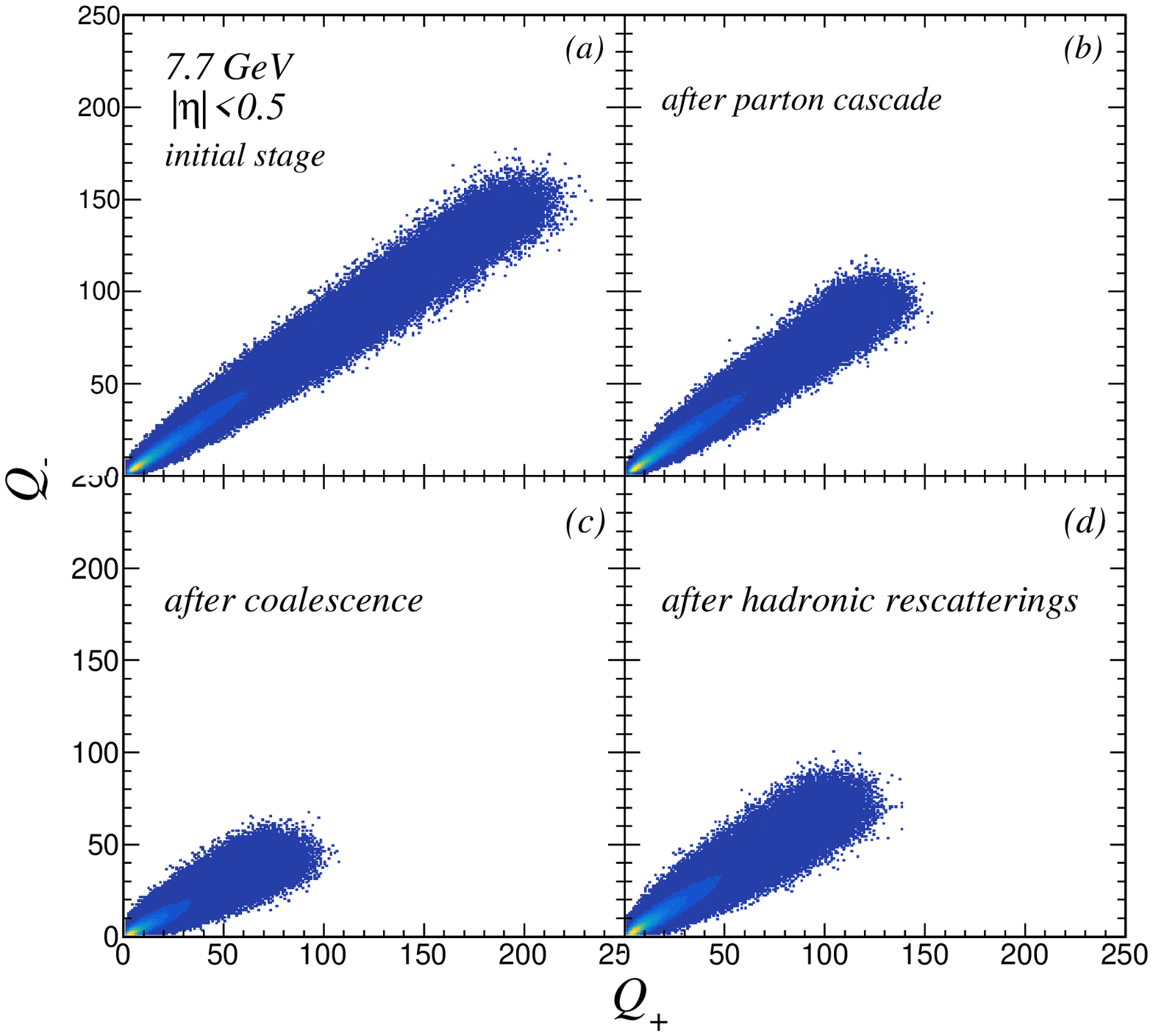}
\end{minipage}
\begin{minipage}[t]{80mm}
\includegraphics[width=80mm]{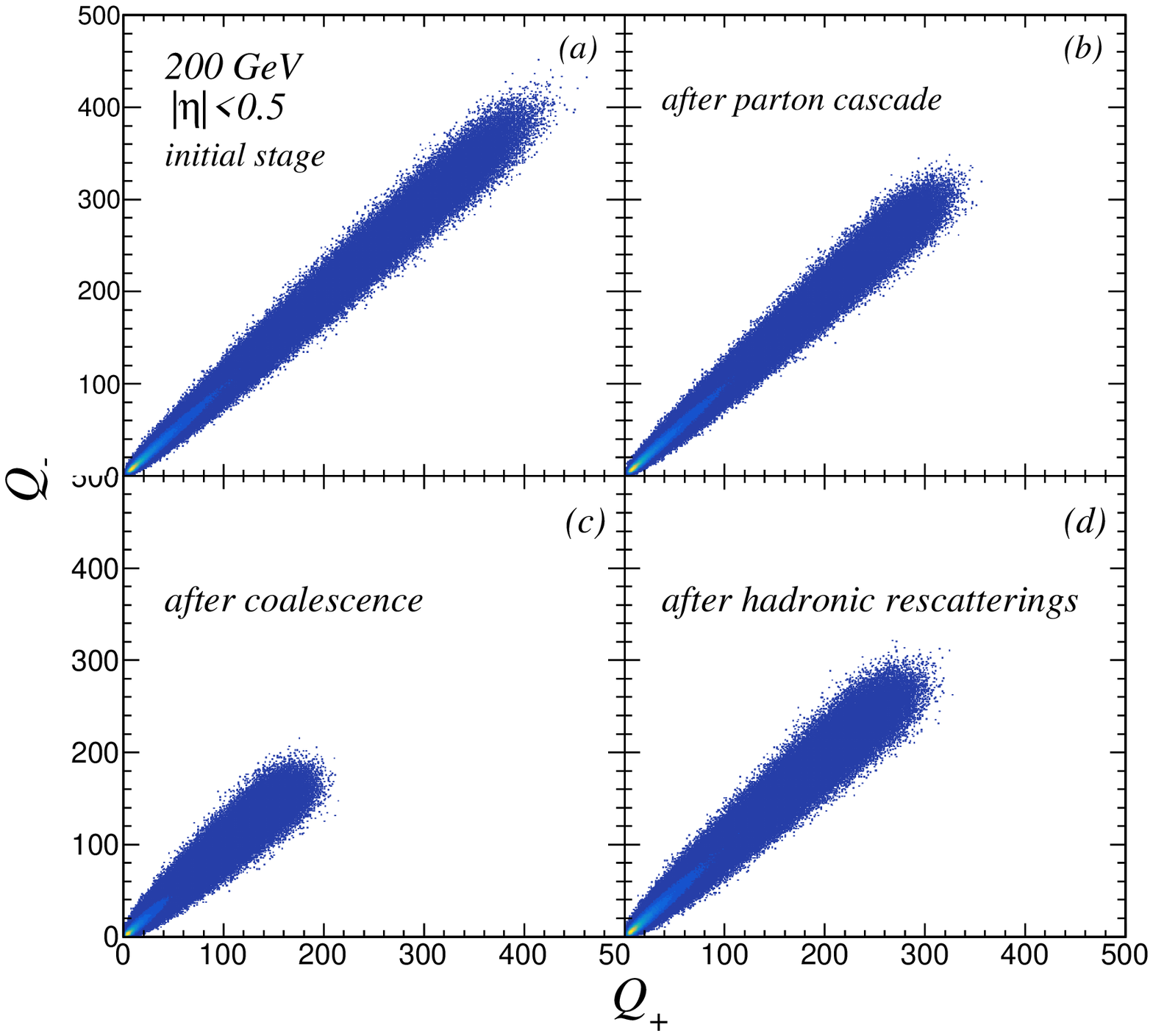}
\end{minipage}
\caption{\footnotesize(Color online) $Q_{-}$ vs $Q_{+}$ at different evolution stages of Au+Au collisions (minus bias) at 7.7 and 200 GeV.} \label{ChargeDis}
\end{figure*}
%%%%%%%%%%%%%%%%%%%%%%%%%%%%%%%%%%%%%%%%%%%%%%%%%%%%%%%%%%%%%%%%%%%%%%%%%%%%%%%
To directly illustrate the two-particle correlation between the multiplicities of positively and negatively charged particles, figure~\ref{ChargeDis} presents the AMPT results on $Q_{-}$ vs $Q_{+}$ at four different stages of the Au+Au collisions (minus bias) at 7.7 and 200 GeV. It can be easily observed that the initial correlation is weakened by parton cascade and then reduced by coalescence, however, it is enhanced by hadronic rescatterings. As indicated in Eq.~(\ref{C2netq}), the change of two-particle correlation could lead to the dynamical evolutions of $\sigma^{2}/M$ for the net charge distributions in relativistic heavy-ion collisions. 

%%%%%%%%%%%%%%%%%%%%%%%%%  StddmeanStage  %%%%%%%%%%%%%%%%%%%%%%%%%%%%%%%%%%%%%%%%%%%%
\begin{figure*}[htb]
\begin{minipage}[t]{80mm}
 \includegraphics[width=80mm]{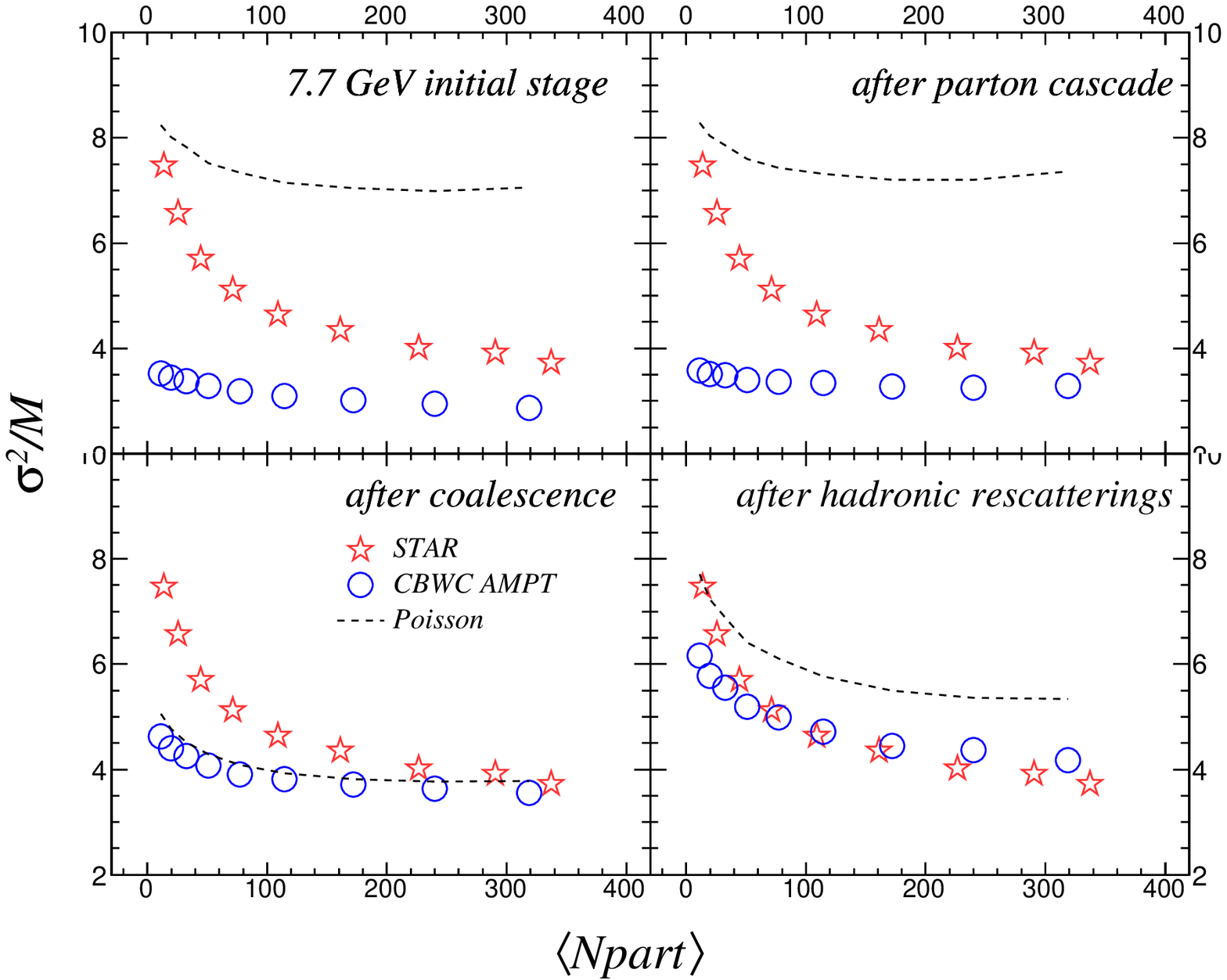}
\end{minipage}
\begin{minipage}[t]{80mm}
\includegraphics[width=80mm]{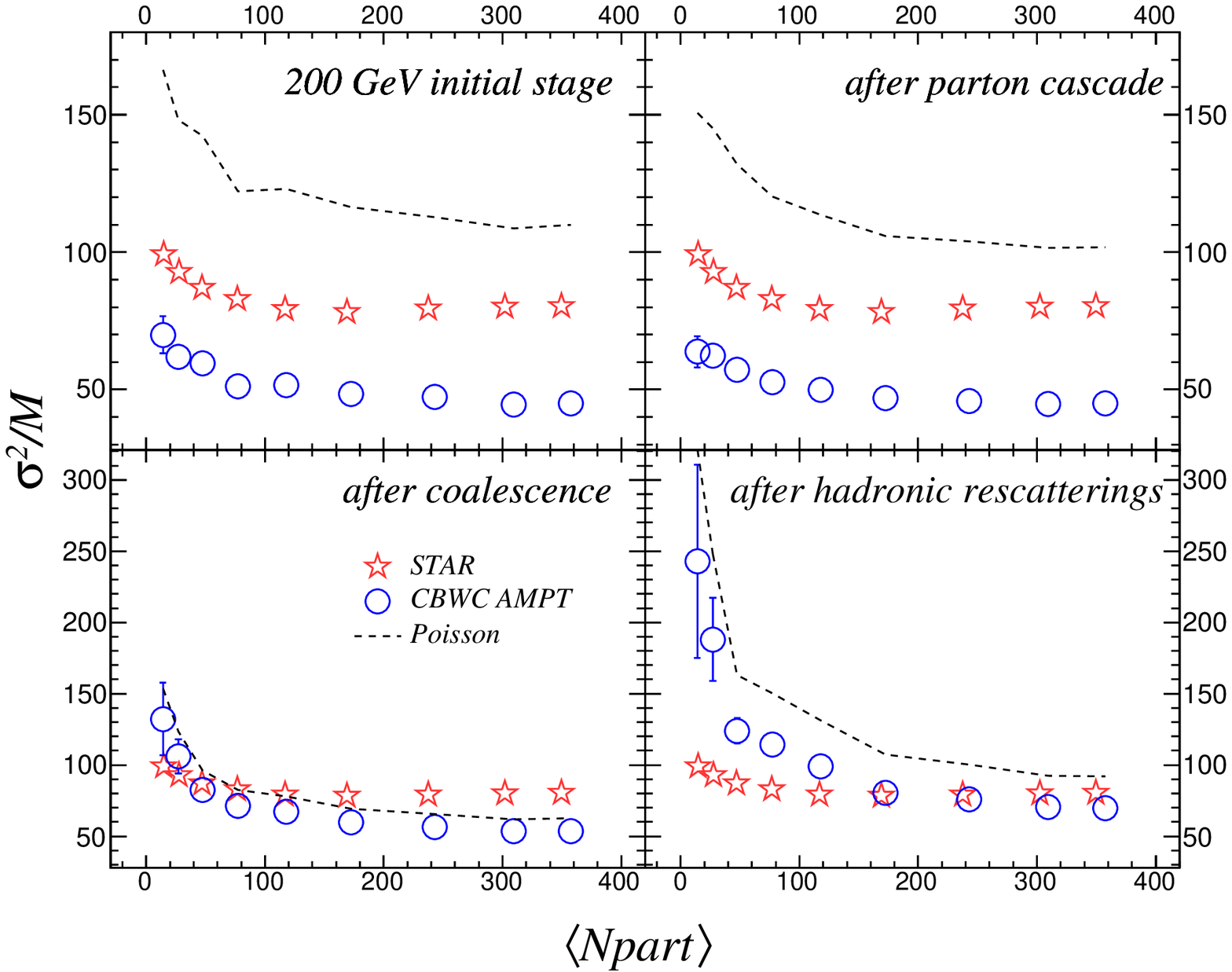}
\end{minipage}
\caption{\footnotesize(Color online) Centrality dependence of moment product $\sigma^{2}/M$ at different evolution stages of Au+Au collisions at 7.7 and 200 GeV, where the STAR data are obtained from Ref.~\cite{Adamczyk:2014fia}.} 
\label{StddmeanStage}
\end{figure*}
%%%%%%%%%%%%%%%%%%%%%%%%%%%%%%%%%%%%%%%%%%%%%%%%%%%%%%%%%%%%%%%%%%%%%%%%%%%%%%%

Figure~\ref{StddmeanStage} presents the centrality dependence of the moment product $\sigma^{2}/M$ at four different evolution stages of Au+Au collisions at 7.7 and 200 GeV, where both experimental results and Poisson expectations are also shown. Note that because the experimental results are obtained from the measured final hadrons, it is only meaningful to compare them with the AMPT results after hadronic rescatterings. The AMPT model can properly describe the data for Au+Au collisions at 7.7 GeV, as well as the mid-central and central bins of Au+Au collisions at 200 GeV; however, it overestimates the data for peripheral bins of Au+Au collisions at 200 GeV. We also observe that $\sigma^{2}/M$ gradually increases from initial stage to final stage. When comparing with the Poisson expectation, it can be observed that the AMPT result is always lower than the corresponding Poisson expectation, which indicates that there is a positive correlation between positively and negatively charged particles, i.e. the two-particle correlation on the right hand of Eq.~(\ref{C2netq}) is always positive. This indicates that two oppositely charged particles tend to appear or disappear together. Since the difference changes gradually stage by stage, which means that the positive correlation develops during the dynamical evolution of heavy-ion collisions. The possible sources of the positive two-particle correlation in the AMPT model are presented as follows.  The positive correlation at the initial stage may originate from the melting process of strings during which the excited strings are decomposed into their constituent quarks and antiquarks with opposite electric charges. After parton cascade, the correlation can be modified because parton collisions can alter the kinematics of partons. However, because the degree of freedom is altered from the parton to hadron after the hadronization of coalescence, the correlation is found to be significantly modified. The difference has almost disappeared, which implies that there is hardly any correlation after the hadronization. The final hadronic rescatterings can not only modify the correlation via hadron rescatterings, similar to parton cascade, but also provide additional correlations owing to resonance decays that usually transform a mother hadron into two daughter hadrons with opposite charges. It appears that the residue correlation in the experimental observable of $\sigma^{2}/M$  is mainly produced in the process of hadronic rescatterings.

%%%%%%%%%%%%%%%%%%%%%%%%%  SkewstddStage  %%%%%%%%%%%%%%%%%%%%%%%%%%%%%%%%%%%%%%%%%%%%
\begin{figure*}[htb]
\begin{minipage}[t]{80mm}
 \includegraphics[width=80mm]{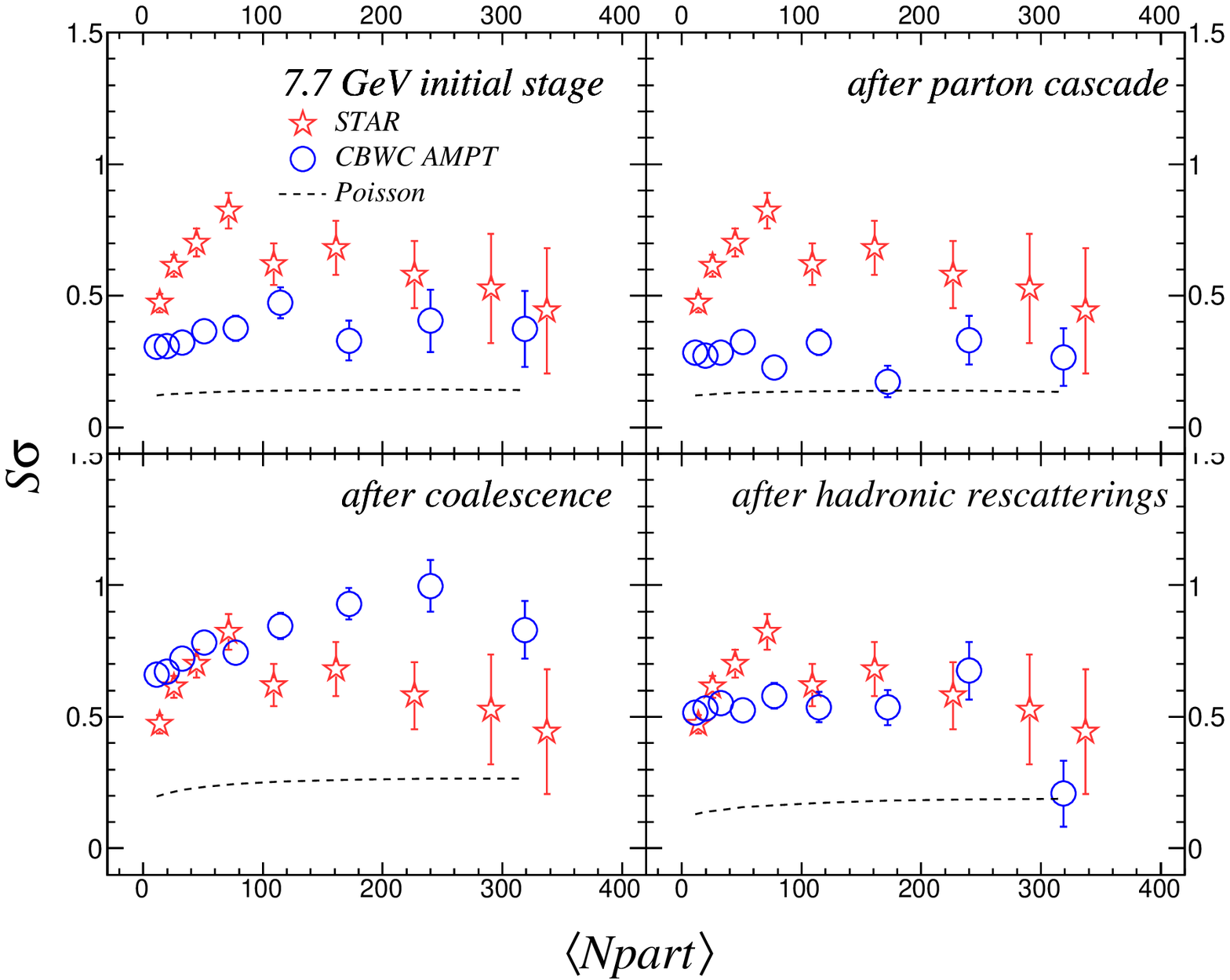}
\end{minipage}
\begin{minipage}[t]{80mm}
\includegraphics[width=80mm]{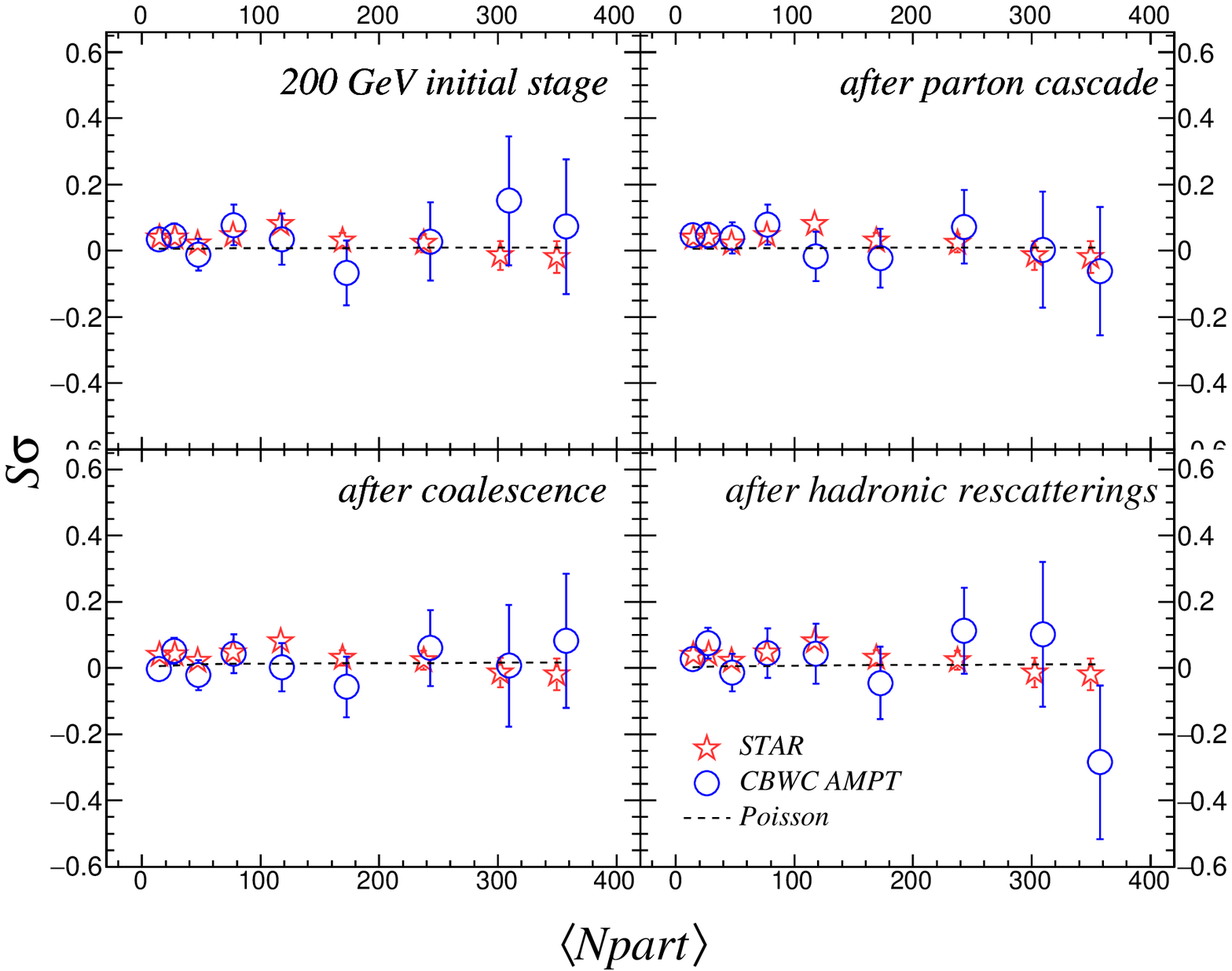}
\end{minipage}
\caption{\footnotesize(Color online) Centrality dependence of moment product $S\sigma$ at different evolution stages of Au+Au collisions at 7.7 and 200 GeV, where the STAR data are obtained from Ref.~\cite{Adamczyk:2014fia}.} \label{SkewstddStage}
\end{figure*}
%%%%%%%%%%%%%%%%%%%%%%%%%%%%%%%%%%%%%%%%%%%%%%%%%%%%%%%%%%%%%%%%%%%%%%%%%%%%%%%

Figure~\ref{SkewstddStage} presents the centrality dependence of moment product $S\sigma$ at four different evolution stages in Au+Au collisions at 7.7 and 200 GeV.  The AMPT model can well describe the experimental data for both energies.  For Au+Au collisions at 200 GeV, the $S\sigma$ at each stages tends to remain approximately zero. However, we observe there is a more significant stage evolution of $S\sigma$ in Au+Au collisions at 7.7 GeV, which indicates that the $S\sigma$ decreases slightly after parton cascade, but increases owing to the hadronization of coalescence, and decreases again after hadronic rescatterings. Because the AMPT results are greater than the Poisson expectations, the total correlation from  the third cumulant $K_3$ should be negative in Au+Au collisions at 7.7 GeV. From Eq.~(\ref{C3netq}), it can be observed that if positively and negatively charged particles are symmetrically related, the total correlation from $K_3$ must be zero. This feature could explain why the deviation from the Poisson expectation is very small and the stage evolution of $S\sigma$ is not quite significant in Au+Au collisions at 200 GeV, because it should be expected that the system of Au+Au collisions at 200 GeV is more electrically symmetric (with a lower electric chemical potential) than that of Au+Au collisions at 7.7 GeV. In contrast, because there are more positively charged particles than negatively charged particles in the system of Au+Au collisions at 7.7 GeV, the forementioned positive two-particle correlation could have a negative impact on the overall correlation of $K_3$, as indicated by the last two lines of Eq.~(\ref{C3netq}).

%%%%%%%%%%%%%%%%%%%%%%%%%  KurtstddStage  %%%%%%%%%%%%%%%%%%%%%%%%%%%%%%%%%%%%%%%%%%%%
\begin{figure*}[htb]
\begin{minipage}[t]{80mm}
 \includegraphics[width=80mm]{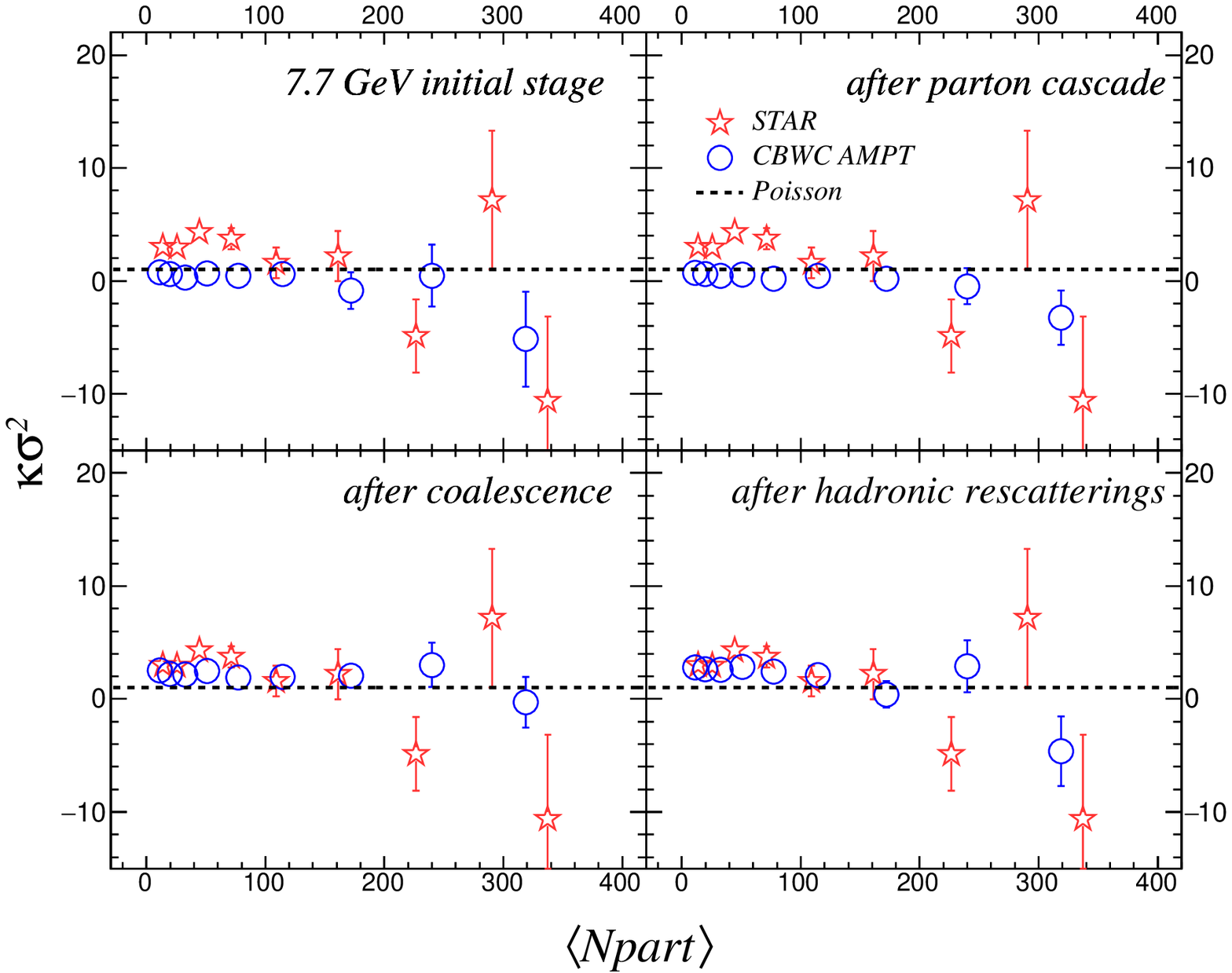}
\end{minipage}
\begin{minipage}[t]{80mm}
\includegraphics[width=80mm]{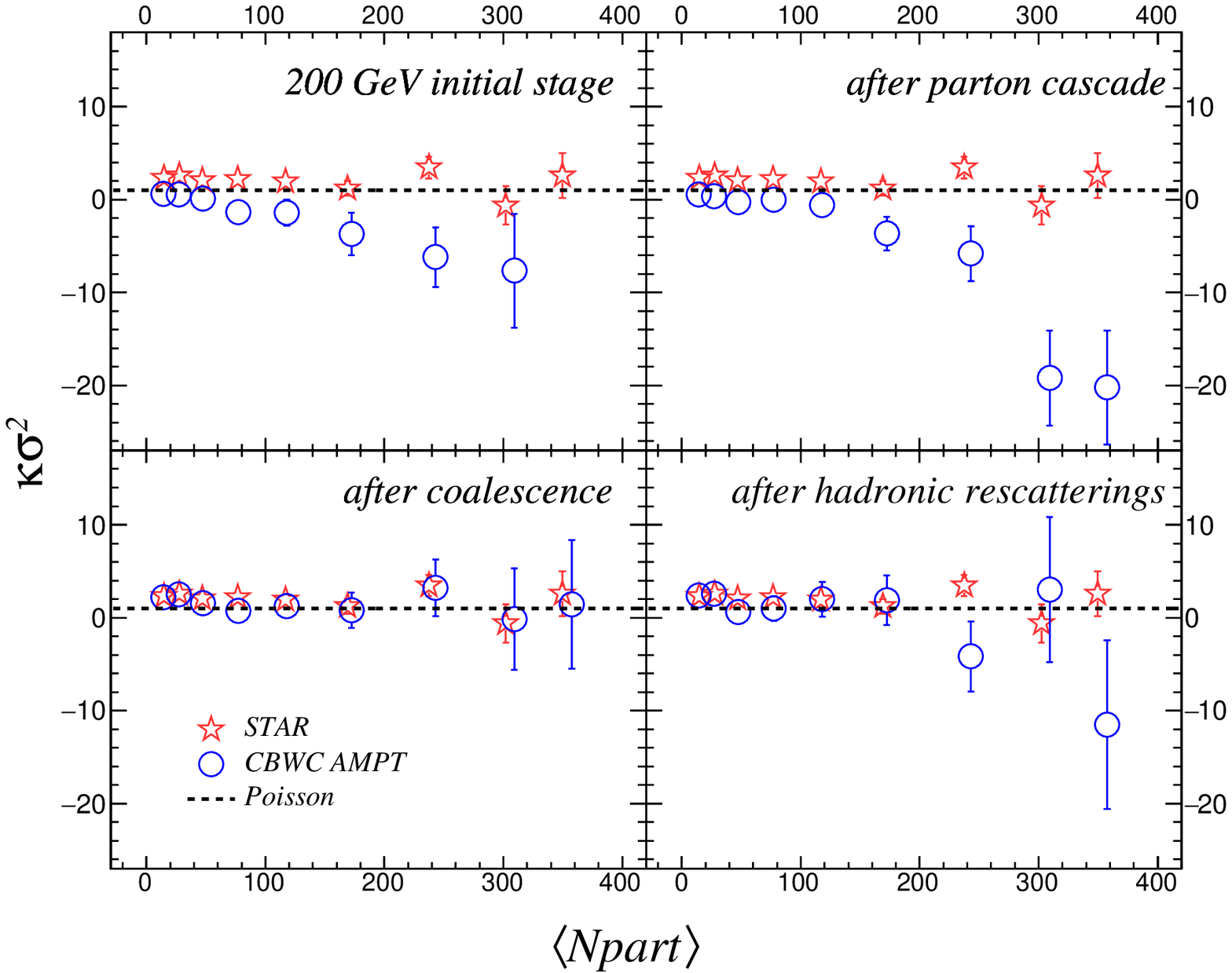}
\end{minipage}
\caption{\footnotesize(Color online) Centrality dependence of of moment product $\kappa\sigma^{2}$ at different evolution stages of Au+Au collisions at 7.7 GeV and 200 GeV, where the STAR data are obtained from Ref.~\cite{Adamczyk:2014fia}.} 
\label{KurtstddStage}
\end{figure*}
%%%%%%%%%%%%%%%%%%%%%%%%%%%%%%%%%%%%%%%%%%%%%%%%%%%%%%%%%%%%%%%%%%%%%%%%%%%%%%%
Figure~\ref{KurtstddStage} presents the centrality dependence of the moment product $\kappa\sigma^{2}$ at four different evolution stages in Au+Au collisions at 7.7 GeV and 200 GeV.  It can be observed that the AMPT results after hadronic rescatterings can basically describe the experimental data of two energies, which is slightly higher than the Poisson expectation of unity. From the AMPT results at different stages, they appear basically consistent with the Poisson expectation, except those before and after parton cascade for central centrality bins which could be caused by the two-particle component of four-particle correlation. According to Eq.~(\ref{C4netq}), the sources of the correlation in the fourth cumulant $K_4$ are significantly complicated, which may be caused by two-particle, three-particle or four-particle correlations. It is believed that the four-particle correlation of net-proton multiplicity distribution is very sensitive to QCD critical fluctuations~\cite{Abdallah:2021fzj,Bzdak:2019pkr,Bzdak:2016sxg}. However, the presence or absence of four-particle correlation in net-charge multiplicity distribution remains an open question that requires more detailed investigations in the future.

It is important to mention that the hadron resonance gas (HRG) model~\cite{BraunMunzinger:2003zd} has indicated that resonance decays could play an important role in influencing these moment products, which enhance $\sigma^{2}/M$ at high colliding energies, suppress $S\sigma$ at low colliding energies, and almost have no impact on $\kappa\sigma^{2}$ for the entire range of colliding energy~\cite{Garg:2013ata,Mishra:2016qyj}. We infer that our results are basically consistent with the observed effect from the HRG model, when compared with the results from the last two stages. However, the extent of impact could be different, as our model includes not only resonance decays but also hadronic elastic and inelastic reactions during the hadronic phase evolution. Therefore, it is interesting to separately study the effects from resonance decays and hadronic reactions in the future.

\section{Summary}
\label{sec:summary}

Using a string melting version of a multi-phase transport model, moments and moment products of net-charge multiplicity distributions in Au+Au collisions at seven different energies of the BES program at RHIC have been studied. It was determined that the AMPT model can basically describe the measured centrality dependences at different energies. After considering the impacts of the CBWE and CRE, we determined that they can affect the values of moments especially for more peripheral collisions at lower energies. The measured energy dependences of the moment products were also optimally reproduced by the AMPT model, and no non-monotonic energy dependence was observed. Through the stage evolution of the moment products, we determined that the moment products develop during the dynamical evolution of heavy-ion collisions. The deviation from the Poisson expectation indicates the existence of the positive two-particle correlation between positively and negatively charged particles, which may be caused by different dynamical processes at different evolution stages. Therefore, to adopt the moments and moment products of net-charge multiplicity distributions in determining the QCD critical point of relativistic heavy-ion collisions, it is important to consider the dynamical evolution.

\section*{ACKNOWLEDGMENTS}

G.L.M. thanks the Department of Physics at East Carolina University for its hospitality, where part of the work was performed. This work is supported in part by the Strategic Priority Research Program of Chinese Academy of Sciences under Grant No. XDB34030000, the National Natural Science Foundation of China under Contracts No. 11890710, No. 11890714, No. 11835002, and No. 11961131011, and the Guangdong Major Project of Basic and Applied Basic Research under Grant No. 2020B0301030008.

\end{document}